%% file: main.tex
\documentclass[conference]{IEEEtran}

\usepackage{booktabs} 
\usepackage{graphicx}
\usepackage{epstopdf}
\usepackage{makecell}
\usepackage[flushleft]{threeparttable}
\usepackage{listings}
\usepackage{xcolor}
\usepackage{balance}
\usepackage{amsmath, amssymb}
\usepackage{tabu}
\usepackage{xspace}

\usepackage{textcomp,booktabs}
\usepackage{colortbl}

\usepackage{marvosym}
\usepackage{algorithm}
\usepackage{algorithmicx}
\usepackage{algpseudocode}

\usepackage[all]{nowidow}

\renewcommand{\paragraph}[1]{\vspace{0.02in}\noindent{\bf{#1}.}}
\newcommand{\sysname}{{\sc DRLgencert}\xspace}

\usepackage[hyphens]{url}
\usepackage[hyphenbreaks]{breakurl}

\definecolor{mygreen}{rgb}{0,0.6,0}
\definecolor{mygray}{rgb}{0.5,0.5,0.5}
\definecolor{mymauve}{rgb}{0.58,0,0.82}
\definecolor{ashgrey}{rgb}{0.7, 0.75, 0.71}
\definecolor{mygrey}{rgb}{0.85, 0.85, 0.85}

\begin{document}
\title{\sysname: Deep Learning-based Automated Testing of Certificate Verification in SSL/TLS Implementations}

\author{
	\IEEEauthorblockN{Chao Chen\IEEEauthorrefmark{1},  Wenrui Diao\IEEEauthorrefmark{2}{\textsuperscript {\normalsize \Letter}}, Yingpei Zeng\IEEEauthorrefmark{3}, Shanqing Guo\IEEEauthorrefmark{1}{\textsuperscript {\normalsize \Letter}}, and Chengyu Hu\IEEEauthorrefmark{1}}
	
	\IEEEauthorblockA{\IEEEauthorrefmark{1}Shandong University, Jinan, China\\
		Email: 1163307648@mail.sdu.edu.cn, \{guoshanqing, hcy\}@sdu.edu.cn}
	\IEEEauthorblockA{\IEEEauthorrefmark{2}Jinan University, Guangzhou, China\\
		Email: diaowenrui@link.cuhk.edu.hk}
	\IEEEauthorblockA{\IEEEauthorrefmark{3}China Mobile (Hangzhou) Information Technology Co., Ltd., Hangzhou, China\\
		Email: zengyingpei@cmhi.chinamobile.com}
}

\maketitle
\input{abstract}


\IEEEpeerreviewmaketitle

\input{introduction}
\input{background}
\input{overview}
\input{testing}
\input{network}
\input{evaluation}
\input{relatedwork}
\input{limitation}
\input{conclusion}
\input{acknowledge}

\balance
\bibliographystyle{IEEEtranS}
\bibliography{bibliography}

\end{document}

%% file: abstract.tex
\begin{abstract}
	
The Secure Sockets Layer (SSL) and Transport Layer Security (TLS) protocols are the foundation of network security. The certificate verification in SSL/TLS implementations is vital and may become the ``weak link'' in the whole network ecosystem. In previous works, some research focused on the automated testing of certificate verification, and the main approaches rely on generating massive certificates through randomly combining parts of seed certificates for fuzzing. Although the generated certificates could meet the semantic constraints, the cost is quite heavy, and the performance is limited due to the randomness.

To fill this gap, in this paper, we propose DRLgencert, the first framework on applying deep reinforcement learning to the automated testing of certificate verification in SSL/TLS implementations. DRLgencert accepts ordinary certificates as input and outputs the newly generated certificates which could trigger discrepancies with high efficiency. Benefited by the deep reinforcement learning, when generating certificates, our framework could choose the best next action according to the result of a previous modification, instead of simple random combinations. At the same time, we developed a set of new techniques to support the overall design, like new feature extraction method for X.509 certificates, fine-grained differential testing, and so forth. Also, we implemented a prototype of DRLgencert and carried out a series of real-world experiments. The results show DRLgencert is quite efficient, and we obtained 84,661 discrepancy-triggering certificates from 181,900 certificate seeds, say around 46.5\% effectiveness. Also, we evaluated six popular SSL/TLS implementations, including GnuTLS, MatrixSSL, MbedTLS, NSS, OpenSSL, and wolfSSL. DRLgencert successfully discovered 23 serious certificate verification flaws, and most of them were previously unknown.

\end{abstract}

%% file: introduction.tex
\section{Introduction}
The Transport Layer Security (SSL)~\cite{dierks2008transport} and its predecessor Transport Layer Security (TLS)~\cite{freier2011secure} protocols are the foundation of network security. They are designed to provide security and data integrity assurance for Internet communications. During the SSL/TLS communication, the X.509 certificate is used to authenticate the identity of communicating party. The semantics and syntax of X.509 certificates have many limitations, described semi-formally in dozens of IETF's RFCs (including RFC 2246\cite{dierks1999rfc}, 2527\cite{chokhanirfc2527}, 2818\cite{rescorla2005rfc}, 4346\cite{dierks2006rfc}, 5246\cite{dierks2008rfc}, 5280\cite{cooper2008internet}, 6101\cite{freier2011secure}, and 6125\cite{saint2011representation}). Verifying the validity of a certificate is a complex process that includes verifying each certificate in the certificate chain, checking period of validity, public key, extension, and so forth.

There are many open source SSL/TLS implementations available online, and developers can use these implementations for certificate validation without implementing the code by themselves. However, since the certificate validation is complicated, described in many tedious documents, the authors of SSL/TLS implementations may have their own understandings of how to code the logic. In other words, there is no guarantee that these SSL/TLS implementations can verify each certificate correctly, which may become the ``weak link'' in the whole network ecosystem.

Some researchers\cite{brubaker2014using,chen2015guided} have noticed this issue and tried to achieve the automated testing of certificate verification in SSL/TLS implementations. One of the mainstream approaches is the differential testing. If multiple certificate verification codes give different verification results for the same certificate, it means some of the certificate verification codes may be flawed. In the differential testing, the effectiveness of test results is decided by the quality of input test cases (i.e., modified certificates) directly. Better test cases could discover more design flaws with less time consumption.

In previous attempts, it is common to generate massive certificates (as test cases) through randomly combining parts of seed certificates, like Frankencert~\cite{brubaker2014using} and Mucert~\cite{chen2015guided}. Although the generated certificates could meet the semantic constraints,  the cost is quite heavy, and the performance is limited due to the randomness.  It may generate a large number of unhelpful certificates which cannot trigger any flaw. On the other side, the deep learning technology shows the powerful capability of information mining and has been widely applied in biology, medicine, graphics, and cybersecurity, and so forth. We find the deep learning is suitable for the task of automated testing of certificate verification.


\vspace{2pt}
\noindent
\textbf{Our Approach.} In this paper, we propose DRLgencert (using Deep Reinforcement Learning to generate certificates), the first framework on applying deep reinforcement learning to the automated testing of certificate verification in SSL/TLS implementations. DRLgencert accepts ordinary certificates as input and outputs the newly generated certificates which could trigger discrepancies with high efficiency. Benefited by the deep reinforcement learning, when generating certificates, our framework could choose the best next action according to the result of a previous modification, instead of simple random combinations. At the same time, we developed a set of new techniques to support the overall design, like new feature extraction method for X.509 certificates, fine-grained differential testing module, and so forth.

In our research, we implemented DRLgencert and carried out a series of real-world experiments based on six popular SSL/TLS implementations, including GnuTLS~\cite{gnutls}, MatrixSSL~\cite{matrixssl}, mbedTLS~\cite{mbedtls}, NSS~\cite{NSS}, OpenSSL~\cite{openssl}, and wolfSSL~\cite{wolfssl}. The overall performance analysis shows DRLgencert is quite efficient, and we could obtain 84,661 discrepancy-triggering certificates from 181,900 certificate seeds after the first training episode of DRL network. It means more than 46.5\% generated certificates are effective, which is better than all existing works. Also, DRLgencert successfully discovered 23 serious certificate verification flaws on six implementations, and most of them were previously unknown. For example, we found GnuTLS, NSS, and OpenSSL accept version 1, 2, or 4 certificates, even these v1, v2, v4 certificates have v3 extensions that should only exist in version 3 certificate, which violates the RFC documents~\cite{cooper2008internet}. The reason is that the version testing and extension testing functions are implemented as two independent components in code. We have reported all these flaws to the corresponding vendors, and the assessments are in process.

\vspace{2pt}
\noindent
\textbf{Contributions.} The main contributions of this paper are:

\begin{itemize}
	\item \textit{New framework.} We proposed DRLgencert, the first framework on applying deep reinforcement learning to the automated testing of certificate verification in SSL/TLS implementations. DRLgencert accepts normal certificates as input and outputs the newly generated certificates (as the test cases of differential testing) which could trigger discrepancies with high efficiency.
	
	\item \textit{New techniques.} We developed a set of new techniques to enable the overall design of DRLgencert, including new feature extraction method for X.509 certificates, fine-grained differential testing, and improved certificate modification actions.
	
	\item \textit{Implementation and findings.} We implemented a prototype of DRLgencert and evaluated it through a series of real-world experiments. The overall performance analysis shows DRLgencert is quite efficient, and we obtained 84,661 discrepancy-triggering certificates from 181,900 certificate seeds. Also, DRLgencert successfully discovered 23 serious certificate verification flaws on six popular SSL/TLS implementations, and most of them were previously unknown.

\end{itemize}

\vspace{2pt}
\noindent
\textbf{Roadmap.} The rest of this paper is organized as follows. Section~\ref{sec:background} gives the background of certificate validation and deep reinforcement learning. Section~\ref{section: system-overview} provides the overview design of DRLgencert. Section~\ref{section:Differential Testing} and Section~\ref{section:network} illustrate the differential testing module and deep reinforcement learning network module respectively. The evaluation results are summarized in Section~\ref{sec:evaluation}. Section~\ref{sec:relatedwork} reviews the related work, and finally
Section~\ref{sec:conclusion} concludes this paper.


%% file: background.tex
\section{Background}
\label{sec:background}

\subsection{Certificate Validation}
\label{section:Certificate Validation}

\begin{figure}
	\includegraphics[width=\columnwidth]{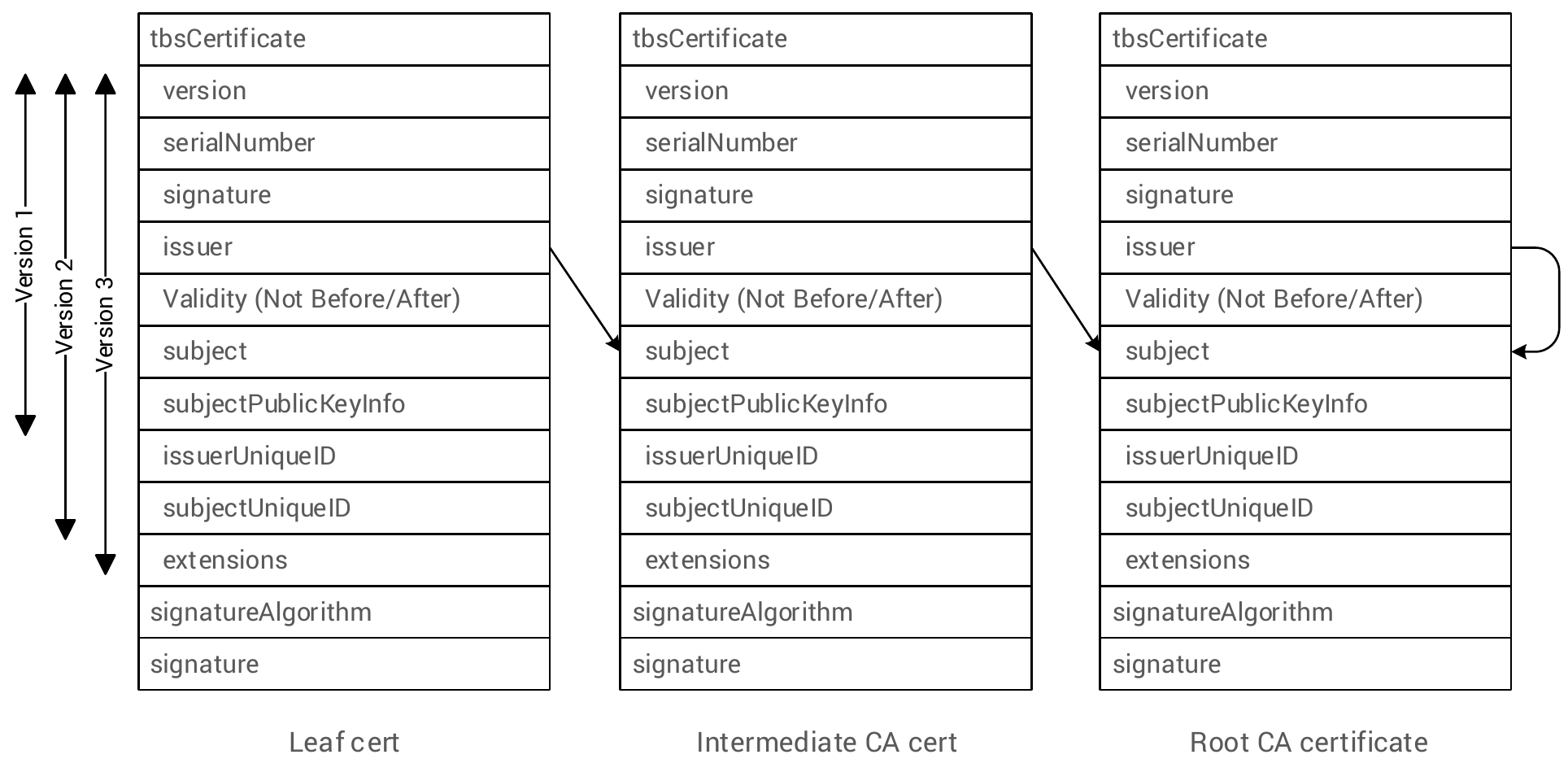}
	\caption{General certificate chain structure.}
	\label{fig:certchain}
\end{figure}

Certificate validation usually requires two inputs: trusted CA certificates and a chain of certificates to be validated. The general structure of a certificate chain is shown in Figure~\ref{fig:certchain}. Starting from the leaf certificate (end-entity certificate), each certificate is issued by a superior certificate until a self-signed CA certificate.

In the verification process, the leaf certificate in the certificate chain is usually verified first to confirm that the content of the certificate is valid, including the validity period and extension content. Then the SSL/TLS client checks whether the certificate is issued by a higher-level certificate in the certificate chain, including verifying the issuer, signature, and extension content. This process continues recursively until the certificate to be verified appears in the trusted certificate set.

\subsection{Deep Reinforcement Learning}
\label{section:Deep Reinforcement Learning}
Deep reinforcement learning is a combination of deep learning and reinforcement learning, which enables robots to learn independently. The initial achievement is the Deep Q Learning algorithm proposed by DeepMind in 2013~\cite{mnih2013playing}.

\begin{figure}[t]
	\includegraphics[width=\columnwidth]{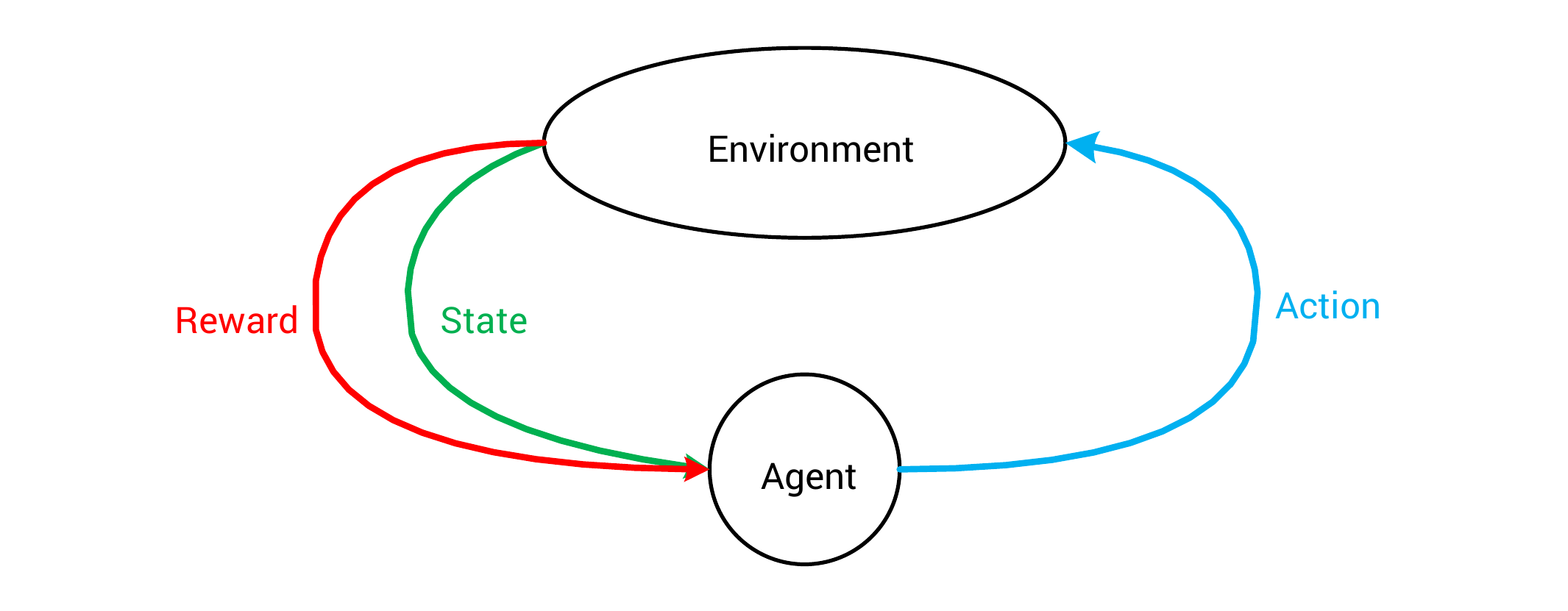}
	\caption{The process of reinforcement learning.}
	\label{fig:rl}
\end{figure}

In the field of artificial intelligence, an agent is used to represent a capable object, such as a robot, an unmanned vehicle, a person, and the like. The problem solved by reinforcement learning is to guide the interaction between the agent and environment. For example, playing a racing game on a computer, the game is the environment, and then we enter actions (keyboard operations) to control. What is observed on the screen is the state of the car and the score. The score is known as the reward, a factor in the reinforcement learning. No matter what kind of task, it contains a series of actions, observations, and feedback value-rewards. The observed information is the state of the agent. When the agent executes the action and interacts with the environment, the environment changes. The reward expresses the degree of good and bad caused by the change. Figure~\ref{fig:rl} illustrates the process of reinforcement learning. Reinforcement learning learns from previous states, actions, rewards to guide the choice of next action.

Combining deep learning with reinforcement learning can handle more complex tasks. Deep reinforcement learning, through continuous training, can get a policy that makes an agent get as much reward as possible in the task.




%% file: overview.tex
\section{Overview of DRLgencert}
\label{section: system-overview}

\begin{figure*}[t]
	\includegraphics[width=\linewidth]{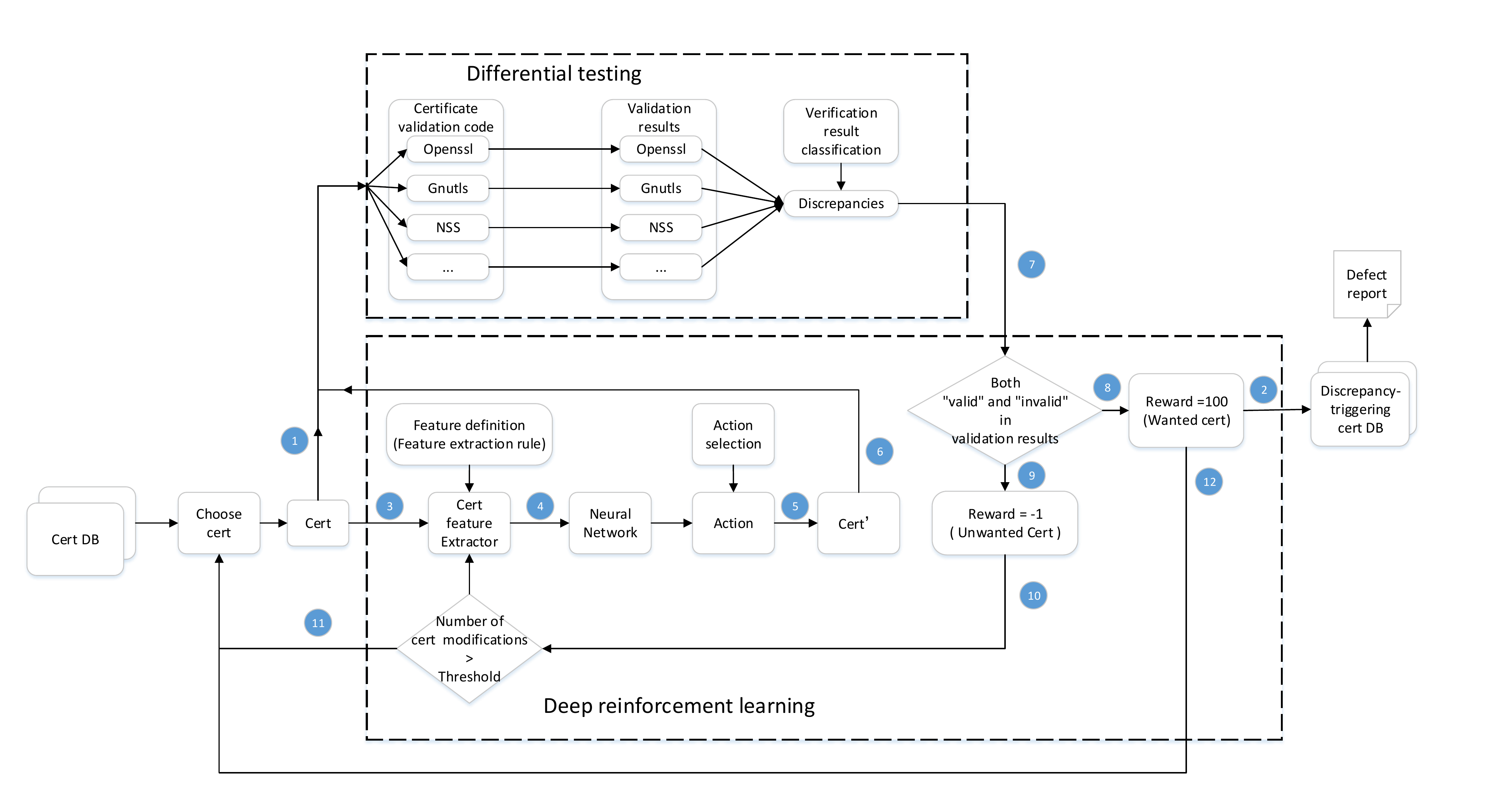}
	\caption{Overview of DRLgencert framework.}
	\label{fig:DRLgencert}
\end{figure*}

In this paper, we propose DRLgencert, the first deep learning-based testing framework for automatic certificate validation in SSL/TLS implementations. DRLgencert accepts normal certificates as input and outputs the newly generated certificates which could trigger discrepancies with high efficiency. As shown in Figure~\ref{fig:DRLgencert}, the DRLgencert framework consists of three main components: \textit{a certificate set}, \textit{the deep reinforcement learning network}, and \textit{the differential testing module} (containing multiple certificate verification programs).


\subsection{Settings}
The reinforcement learning is a cyclic process in which an agent takes actions to change its state and interact with the environment to obtain a reward. In this process, the agent judges the merits and demerits of previous actions according to the reward given by the environment and obtains experience, so that the agent could choose a better behavior in the future same or similar state. In DRLgencert, we have the following settings:

\begin{itemize}
	\item A certificate instance acts as an agent.
	\item Multiple SSL/TLS implementations are used as the environment.
	\item The state is defined as a certificate content feature.
	\item The action is defined as a certificate modification operation.
	\item The reward results from the validation results of multiple SSL/TLS implementations.
\end{itemize}


\subsection{Framework Overview}
DRLgencert starts to run from choosing a certificate instance from the prepared certificate set. Before passing the certificate to the deep learning network, the differential testing (see Section~\ref{section:Differential Testing}) is performed on the original certificate(step1 in fig~\ref{fig:DRLgencert}). If the test result shows that the certificate has reached the goal we want, namely, the certificate triggers a discrepancy(step7,8,2 in fig~\ref{fig:DRLgencert}), there is no need to modify this certificate. Otherwise, we need to extract features from the certificate based on the pre-defined feature extraction scheme(step3 in fig~\ref{fig:DRLgencert}). The definition of certificate feature will be discussed in detail in Section~\ref{section:Feature Definition} later, respectively. The neural network uses the certificate feature as input(step4 in fig~\ref{fig:DRLgencert}), and the output represents the modification actions. Each number of the output represents the value of the corresponding modification action. As mentioned in Section~\ref{section:Deep Reinforcement Learning}, the deep reinforcement learning network learns the feedback coming from the agent interacting with the environment by taking actions. In DRLgencert, the certificate modification actions are the actions taken by the deep reinforcement learning on certificates, and more details will be given in Section~\ref{section:Certificate Modification Action Definition}.


During the training phase, we select the modification action to be adopted based on the $\varepsilon$-Greedy strategy. Each time we choose the most valuable modification action with 90\% probability and randomly select a modification operation with 10\% probability(step5 in fig~\ref{fig:DRLgencert}). As an advantage, we could make sure that most modification operations can achieve the desired effect and also allow some certificate to explore new combinations of modification actions. In the testing phase of the neural network, we no longer need to explore new combinations of actions. Instead, we should guarantee the maximum validity of the modification actions. Therefore, the certificate modification action used during the usage phase is always the most valuable one in all modification actions.


After the certificate is modified, we obtain a new certificate. This newly generated certificate will be inputted to the differential testing for multiple certificate verification programs(step6 in fig~\ref{fig:DRLgencert}). After the differential testing, we obtained the verification result set of each verification program. According to the predefined reward definition scheme, the result is transformed to the corresponding reward(step7,8,9 in fig~\ref{fig:DRLgencert}). The definition of reward will be given in Section~\ref{section:Reward Definition}. In the DRLgencert framework, we set the range of reward for -1 and 100. If reward = 100, it means that the certificate achieves the target we want. It will be collected into our target certificate database(step2 in fig~\ref{fig:DRLgencert}). Its content and verification result set may help us to analyze and discover possible flaws of certificate verification implementations. Then choose a new certificate from database and a new loop begins(step12 in fig~\ref{fig:DRLgencert}). However, if reward = -1, it indicates that, under our settings, this certificate has no value in analyzing certificate verification implementations, and further change actions to the certificate content are required(step10 in fig~\ref{fig:DRLgencert}). This process is described in Algorithm~\ref{algorithm:Modifying}:

\begin{algorithm}[ht]
	\caption{Modifying certificates}
	\label{algorithm:Modifying}
	\begin{algorithmic}[1]
		\Require $certDB, max\_episode, max\_modification$
		\Ensure $target\_certDB$
		\State $episode \gets 0$
		\While{$episode < max\_episode$}
		\For{ $cert$ in $certDB$ }
		\State $modification \gets 0$
		\While {$modification <= max\_modification$}
		\State $action \gets$ \Call{network}{$cert$}
		\State $cert \gets$ \Call{take\_action}{$cert,action$}
		\State $reward \gets$ \Call{differential\_testing}{$cert$}
		\If{$reward == 100$}
		\State add $cert$ to $target\_certDB$
		\State $break$
		\EndIf
		\State $modification \gets modification + 1$
		\EndWhile
		\EndFor
		\State $episode \gets episode + 1$
		\EndWhile
		\State \Return{$target\_certDB$}
	\end{algorithmic}
\end{algorithm}

In our DRLgencert framework, we set $max\_modification$ (in Algorithm~\ref{algorithm:Modifying}) to 9 and allow up to 10 changes per certificate, that is a maximum of 10 variants except for the initial certificate. If this certificate still cannot trigger a discrepancy, we will discard it and select a new one(step11 in fig~\ref{fig:DRLgencert}). In our experiments, we found 95.5\% of the certificates which get reward = 100 were modified less than 6 times. Therefore, if a certificate undoubtedly has the potential to trigger a discrepancy, 10 times is enough to trigger a discrepancy.

Through the above process, DRLgencert helps us quickly obtain a large number of certificates suitable for differential testing, based on the collected certificate data set.

%% file: testing.tex
\section{Differential Testing}
\label{section:Differential Testing}
Differential testing~\cite{mckeeman1998differential} (also known as differential fuzz testing) is a popular software testing technique that attempts to detect errors of a series of similar applications (or different implementations of the same application) by providing the same input and observing their execution differences.

\begin{table}[t]
	\caption{Verification programs and supported modes}
	\centering
	\label{tab:mode}
	\begin{tabular}{llcl}
		\toprule
		\makecell[bl]{\textbf{Program}} & \makecell[bl]{\textbf{Version}} & \makecell[bl]{\textbf{C-S Mode}} & \makecell[bc]{\textbf{File Mode}\\(with verification utility)}\\
		\midrule
		GnuTLS & 3.6.0 & Y & \textbf{Y} (certtool)\\
		MatrixSSL & 3.9.3 & Y & \textbf{Y} (certValidate)\\
		MbedTLS & 2.6.0 & Y & \textbf{Y} (cert\_app)\\
		NSS & 3.28.4 & Y & \textbf{Y} (certutil)\\
		OpenSSL & 1.0.2g & Y & \textbf{Y} (openssl)\\
		wolfSSL & 3.12.2 & \textbf{Y} & /\\
		\bottomrule
	\end{tabular}
	\begin{tablenotes}
		\item Remarks: The mode we used is labeled with bold \textbf{Y}.
	\end{tablenotes}
\end{table}

\vspace{2pt} \noindent
\textbf{Verification Program Setup.} In DRLgencert, the initial certificates and modified certificates are both classified by six verification programs, including GnuTLS~\cite{gnutls}, MatrixSSL~\cite{matrixssl}, mbedTLS~\cite{mbedtls} (formerly PolarSSL), NSS~\cite{NSS}, OpenSSL~\cite{openssl}, and wolfSSL~\cite{wolfssl} (formerly CyaSSL). We used the latest versions of these programs (as listed in Table~\ref{tab:mode}) and deployed them on Ubuntu 16.04.



Most of these verification programs provide two verification methods: C-S mode and file mode. The only exception is wolfSSL which only supports the C-S mode. In the C-S mode, the client establishes a connection with the server, receives the certificate provided by the server, and validates it. Compared with the C-S mode, the file mode is more convenient. The program directly loads trust certificates and the certificate to be verified, then verifies it. In both modes, when they need to verify a certificate, they call the same functions to verify. The verification process is the same for both. However, because in the c-s mode, the program needs to establish a connection between the client and server first, the operation is more tedious, and the connection may fail due to network reasons and the certificate verification section cannot be performed. In order to speed up the verification process and avoid the verification failure caused by the connection issue, we selected the file mode as the certificate verification method with a priority in DRLgencert. If the file mode is not available, we used the C-S mode. Table~\ref{tab:mode} shows the mode we selected in each program.

\vspace{2pt} \noindent
\textbf{Verification Result Processing.} Since each verification program has its own understanding of certificate verification, it leads to different expressions for the same error. Also, the meanings of the same error code from different programs are usually diverse. The granularity of verification results of different programs is not the same, and the verification result sets are also different.

\begin{table*}[t]
	\centering
	\begin{threeparttable}
		\caption{Verification results classification}
		\label{tab:classification}
		\begin{tabular}{|c|c|c|c|c|c|c|c|}
			\toprule
			\textbf{Verification Result}& \textbf{Error Code} & \textbf{GnuTLS} & \textbf{MatrixSSL} & \textbf{MbedTLS} & ~~~\textbf{NSS}~~~ & \textbf{OpenSSL} & \textbf{wolfSSL} \\
			\midrule
			Valid& 1& $\surd$ & $\surd$ & $\surd$ & $\surd$ & $\surd$ & $\surd$ \\
			\hline
			Unknown issuer& -1& $\surd$ & & & ~$\surd$$\ast$ & $\surd$ & $\surd$\\
			\hline
			Validity period error& -2& ~$\surd$$\ast$ & $\surd$ & $\surd$ & ~$\surd$$\ast$ & $\surd$ & $\surd$\\
			\hline
			Parsing error& -3& ~$\surd$$\ast$ & & $\surd$ & ~$\surd$$\ast$ & $\surd$ & \\
			\hline
			Version error& -4& & ~$\surd$$\ast$ & & $\surd$ & & $\surd$ \\
			\hline
			Algorithm error& -5& & ~$\surd$$\ast$ & & & & \\
			\hline
			Signature error& -6& & & ~$\surd$$\ast$ & $\surd$ & $\surd$ & $\surd$\\
			\hline
			Subject/Issuer error& -7& & ~$\surd$$\ast$& $\surd$ & & & \\
			\hline
			Key usage error& -8& & ~$\surd$$\ast$& & $\surd$ & & $\surd$\\
			\hline
			Casic constraints error& -9& & & & & $\surd$ & \\
			\hline
			Unknown critical extension& -10& & $\surd$ & & $\surd$& $\surd$& \\
			\hline
			Chain error& -11& & & & $\surd$& $\surd$& \\
			\hline
			Self sign& -12& & & & & $\surd$& \\
			\hline
			Connection error& -13& & & & & & ~$\surd$$\ast$\\
			\hline
			Other extension error& -14& & ~$\surd$$\ast$ & & & & \\
			\hline
			Other error& -15& & & & & & \\
			\bottomrule
		\end{tabular}
		\begin{tablenotes}
			\item $\surd$ indicates that the verification result has been triggered by the program.
			\item $\ast$ ~indicates that multiple different verification results are grouped into the category.
		\end{tablenotes}
	\end{threeparttable}
\end{table*}

Inspired by the work of Acer et al.~\cite{acer2017wild}, we designed a normalized solution to process the various verification results. In details, we grouped the verification results of each program encountered in our experiment into 16 categories, with a new error code, making it easy to redefine the reward and analyze verification codes. Table~\ref{tab:classification} shows the classification of verification results in our system and the number in the second column (``Error Code'') indicates the new encoding value of verification results. Due to the diverse granularity, sometimes multiple verification results from the same verification program are classified into same result type. If there are any validation results that have not been encountered in previous experiments, they are uniformly grouped into the type of ``\textit{other error}'' and re-classified into an appropriate result category in next experiment through the updated classification policies.

\begin{figure}[t]
	\centering
	\includegraphics[width=\columnwidth]{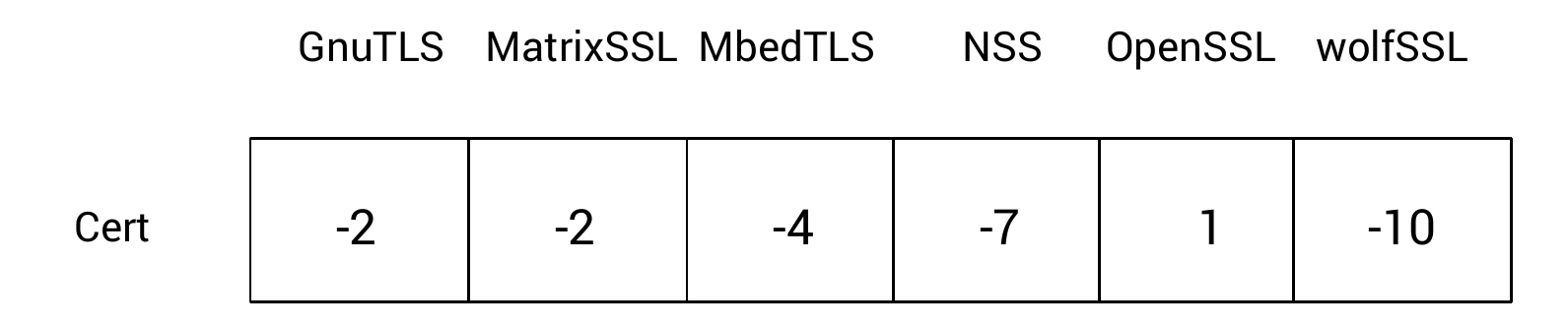}
	\caption{Example of result encoding.}
	\label{fig:encoding}
\end{figure}

Each verification program has its own understanding of the logic of certificate verification. Therefore, different programs may give different verification results for the same certificate, even giving the opposite results. If a certificate is given different verification results (i.e., rejected for different reasons), it means that the validation logic of the involved verification programs is different, and the logic of some programs may be flawed. As shown in the example of Figure~\ref{fig:encoding}, OpenSSL accepts the certificate, while other programs report different rejection reasons respectively.

To our framework, this certificate sample has the value of analysis. In DRLgencert, if there exist both acceptance and rejection in the verification results of a certificate, we define it as a discrepancy and record this certificate and its verification results for later analysis.

%% file: network.tex
\section{Deep Reinforcement Learning Network}
\label{section:network}


The core part of DRLgencert is the deep reinforcement learning network, which is used to generate a large number of certificates suitable for differential testing.

Compared with other popular deep learning algorithms (e.g., CNN and LSTM), deep reinforcement learning is unsupervised, can learn from histories, and provide the optimal choice. The most famous deployment case is AlphaGo~\cite{AlphaGo}, every time it could select the optimal strategy based on the current situation. To be specific, we consider: sometimes the program is very sensitive to the changes in input content, just like the existence of adversarial examples~\cite{goodfellow2014explaining} in image recognition field. That is, the image recognition program can correctly recognize a dog's photo, but if you make minor changes to the photos, the human cannot distinguish between the difference before and after the two photos. However, the image recognition program may identify the modified image as other objects, such as cars, just because of multiple subtle superimposed changes on several certain pixels. To certificates, we think that the certificate also has such nature that the programs is very sensitive to content changes in certificates even the change is very small. It is unlikely that a one-time modification makes a certificate get a different verify result, so the combination of multiple modifications is a better choice. Therefore, we chose deep reinforcement learning, as it can help us choose the best next action according to the result of a previous modification.

\begin{table*}[t]
	\centering
	\caption{Network configuration parameters}
	\label{tab:parameters}
	\begin{tabular}{cccc}
		\toprule
		\makecell[bl]{\textbf{Layer}} & \makecell[bl]{\textbf{Input Dimension}} & \makecell[bl]{\textbf{Output Dimension}} & \makecell[bl]{\textbf{Activation Function}}\\
		\midrule
		Layer\_0: fully connected layer & 101 & 100 & ReLU\\
		Layer\_1: fully connected layer & 100 & 100 & ReLU\\
		Layer\_2: fully connected layer & 100 & 86  & /\\
		\bottomrule
	\end{tabular}
\end{table*}

Based on the practical testing and our empirical knowledge, we deployed a three-layer neural network for DRLgencert. The concrete configuration parameters are listed in Table~\ref{tab:parameters}.

\begin{figure*}[t]
	\begin{equation}
	\label{tab:loss function}
	loss(cert) =
	\begin{cases}
	reward - \max(network(cert))       & \mbox{modifications are finished}  \\
	reward + \gamma\ast(\max(network(cert'))) - \max(network(cert))   &\mbox{modifications are unfinished}  \\
	\end{cases}
	\end{equation}
\end{figure*}

According to the classic definition of  Mnih et al.~\cite{mnih2013playing}, the loss function in DRLgencert is defined as Equation~\ref{tab:loss function}, in which:

\begin{itemize}
	\item $reward$: After $cert$ being changed to $cert'$, $reward$ is gotten from differential testing. 
	\item $\max$$(network(cert))$: Using a $cert$ as input, it is the max value in output.
	\item $\max$$(network(cert'))$: Using $cert'$ as input, it is the max value in output.
	\item $\gamma$ : It is a constant, between 0 and 1. We set it to 0.9 in our experiment.
\end{itemize}

As shown in Figure~\ref{fig:DRLgencert}, the neural network uses the certificate feature as input, and the output represents the modification actions. The reward (from the differential testing result) determines whether further actions are required. In the following sub-sections, we will discuss the designs of feature, modification action, and reward respectively.

\subsection{Feature Extraction}
\label{section:Feature Definition}
Before making a certificate change, we need to characterize the certificate entity and use a vector to represent the content of the certificate. We refer to the feature extraction scheme of Dong et al.~\cite{zheng} and design a more accurate feature extraction scheme.




The certificate consists of a number of different content components. The value type of some parts is the number, others are string or bool. In the DRLgencert framework, we extract the relevant features from the \textit{version}, \textit{issuer}, \textit{subject}, \textit{notafter}, \textit{notbefore}, \textit{public key length}, \textit{signature algorithm}, and \textit{extensions} of the certificate. A total of 101 numbers are used as the feature of a certificate, as the following example.

\vspace{3pt}
\noindent
\fbox{%
	\parbox{0.97\columnwidth}{%
\texttt{[2, 0, 1, 2, 1, -1, 1, 0, 0, 0, 0, 0, 0, 0, 0, 0, 0, 0, 0, 0, 0, 0, 0, 0, 0, 0, 0, 0, 0, 0, 0, 0, 0, 0, 1, 0, 0, 0, 0, 0, 0, 0, 0, 0, 0, 0, 2, 0, 1, 0, 0, 0, 3, 0, 2, 0, 0, 0, 1, 0, 0, 0, 0, 2, 0, 0, 0, 0, 0, 0, 0, 0, 0, 0, 0, 0, 0, 0, 1, 1, 0, 1, 0, 0, 0, 0, 0, 0, 3, 0, 0, 0, 0, 0, 0, 0, 0, 0, 0, 0, 2]}
	}%
}
\vspace{2pt}

\textit{Version.} The version of the current mainstream certificate is 3. Version 1 and version 2 are also possible, but we can not limit version to [1, 2, 3] as we are not sure whether other versions of certificates exist and someone may deliberately make other versions of certificates. Therefore, DRLgencert directly uses the certificate version value as a feature.

\textit{Issuer and subject.} In the certificate verification process, generally, only the issuer of the certificate and the subject of the superior certificate are checked to judge whether they are the same, without regard for the meaning indicated by the values of subject and issuer. However, it does not exclude that some processes of certificates verification codes are rather special, so we extract the value of the country of subject and issuer as certificate features, with number label to represent.

\textit{Notafter and notbefore.} The validity period of certificate is compared with the current time, so we use the comparison result of the validity period and the current time as a feature. -1 means notbefore or notafter is before the current time. 0 means equal and 1 means they are after the current time.

\textit{Public key length.} Public key length extraction scheme is similar to the processing of version. Its value is a numeric type, and cannot be guaranteed to ensure the value of public key length is in a certain range, so public key length uses its value as a feature.

\textit{Signature algorithm.} Collect signature algorithm types that appear in the certificate set. Different numbers are used to label different signature algorithms.

\textit{Extensions.} Extensions are optional content for certificates. The existence and type of extensions in a certificate are not fixed, and users can customize the types of their own extensions. Each extension must contain three contents: oid, iscritical, value. In DRLgencert system, we use three ways to extract extension features. According to the number of different types of extensions in collected certificate set and the difficulty of getting the content feature of a certain extension type, we choose different extraction methods for different kinds of extensions.

\begin{enumerate}
	\item Get feature according to the extension's iscritical and value.
	\item Get feature according to the extension's iscritical and isexist.
	\item Ignore the extension.
\end{enumerate}

Finally, in DRLgencert, we use an array of length 93 to represent extensions.

\subsection{Certificate Modification Action}
\label{section:Certificate Modification Action Definition}
In our settings, if a certificate cannot trigger a discrepancy in the differential testing, namely, there is no difference between verification results of tested programs, we think the certificate and its validation results have no value for further analysis for validation codes. Then we will modify the content of the certificate, expecting the modified certificate will have the value we want. In DRLgencert, we designed a total of 86 modification actions, such as changing 'Version', changing 'Not before', deleting a certain extension. Each action modifies a part of the certificate and does not destroy its semantic constraints. At the same time, the modification operation will try to change the feature value of the certificate. In brief, in order to change content and feature value of a certificate, we designed these modification operations. We set a maximum of nine times of changes per certificate.


\subsection{Reward}
\label{section:Reward Definition}
In the reinforcement learning, after an agent takes action to change its state, the environment will feedback a corresponding reward to show the pros and cons of the result generated by the action. In DRLgencert, the reward is obtained from the discrepancy's encoding. As described in Section~\ref{section:Differential Testing}, we reclassify the verification results of each validation program and encode them with new error codes, so we use a sequence of error codes to represent a discrepancy as its encoding.


In the preliminary experiment, we defined the reward as: assume that before the modification, the number of categories of cert$_1$'s verification results in 6 tested programs is $c_1$. After the modification, cert$_1$ becomes cert$_2$. The number of categories of cert$_2$ verification results is $c_2$. Then we define reward as ($c_2$-$c_1$).

If the reward is greater than 0, or $c_2$ is equal to 6 (the number of verification programs), DRLgencert stops the process of modification and starts to select another certificate from the certificate data set. Because that if the reward is greater than 0, it means there are more types of verification results we achieve. The nature of discrepancy is clarified. DRLgencert helps to generate a discrepancy or make the discrepancy more distinctive. If reward = 6, then there is no need to make more changes. It has reached the maximum value.


However, when analyzing the generated discrepancies, we found that because of the limited manual efforts, this design is not suitable as it generates too many low-value discrepancies. Therefore, we redefined the reward. The new definition is: if acceptance and rejection both exist in the validation result, that is, there are 1 and non-1 in the validation result encoding, the reward is defined as 100. Otherwise, the reward is -1. During the training process, the neural network will be trained to get reward as large as possible. Therefore, essentially, we only need to ensure that the reward value used to mark discrepancy is bigger than that used to mark non-discrepancy. In order to make neural network distinguish between discrepancy and non-discrepancy more distinctly, we expanded the gap between their rewards, empirically set to 100 and -1. When the reward is equal to 100, the certificate's modification process is over, as shown in Figure~\ref{fig:DRLgencert}.

Comparing the effects of the two reward definitions, although the first strategy eventually obtains more discrepancies as defined in previous work, not every discrepancy has high analytical value as many discrepancies are encoded with different rejection reports. The program that accepts a certificate is highly likely to have flaws when all the other programs reject the same certificate. Therefore, from the perspective of differential testing, the discrepancies containing both acceptance and rejection have more value for analysis than those only containing different rejection reports. The second definition helps us filter out these low-value discrepancies.

Since we used manual methods to analyze code based on discrepancies statically, the workload of code analysis is a key factor to consider in system design. If there are enough manual efforts, the first reward definition will be better for finding more flaws as the discrepancies consisting of different rejection reports may also contain hidden flaws. In our case, DRLgencert adopts the second reward definition, spending less time on more valuable discrepancies.

%% file: evaluation.tex
\section{Results}
\label{sec:evaluation}

We implemented DRLgencert and carried out a series of experiments to demonstrate its effectiveness.

\subsection{Certificate Collection}
\label{section:Certificate Collection}
In our experiment, we used ZMap~\cite{durumeric2013zmap} and OpenSSL~\cite{openssl} to collect certificates as seeds. With ZMap, we obtained 300,000 IP addresses that open the 443 port. Then, using OpenSSL s\_client to connect to these IPs, we collected 181,900 certificate samples. One of the failure reasons for getting a certificate is that some IPs opened the 443 port but did not provide their certificates.



\subsection{Certificate Generation}
Table~\ref{tab:discrepancy} shows the types of discrepancies generated by DRLgencert in the experiment. Discrepancies are encoded in the format shown in Figure~\ref{fig:encoding}, and each type of discrepancy contains both acceptance and rejection (negative code are all considered as rejection).

\begin{table*}
	\centering
	\footnotesize
  \caption{Discrepancies generated in the first training episode: Type}
  \label{tab:discrepancy}
  \resizebox{\textwidth}{50mm}{
  \begin{threeparttable}
  \begin{tabular}{ccccccc|ccccccc}
    \toprule
    Sequence number & GnuTLS & MatrixSSL & MbedTLS & NSS & OpenSSL & wolfSSL & Sequence number & GnuTLS & MatrixSSL & MbedTLS & NSS & OpenSSL & wolfSSL\\
    \midrule
    1 & 1 & -14 & -6 & 1 & 1 & 1 & 32 & -3 & -14 & -6 & -1 & -1 & 1 \\
    2 & -1 & -14 & -6 & -4 & -1 & 1 & 32 & -3 & -14 & -6 & -1 & -1 & 1 \\
    3 & 1 & -15 & -6 & 1 & 1 & 1 & 34 & 1 & -15 & -3 & -4 & 1 & -4 \\
    4 & -1 & -15 & -3 & 1 & -1 & -4 & 35 & 1 & -10 & -3 & -4 & -10 & -13 \\
    5 & -3 & -15 & -3 & 1 & 1 & -4 & 36 & 1 & -14 & -3 & 1 & -1 & -4 \\
    6 & -3 & -15 & -3 & -4 & 1 & -4 & 37 & 1 & -15 & -3 & 1 & -10 & 1 \\
    7 & -1 & -15 & -3 & -10 & -1 & 1 & 38 & 1 & -14 & -3 & 1 & 1 & -13 \\
    8 & -3 & -14 & -3 & -4 & -1 & 1 & 39 & -3 & -15 & -6 & -4 & -1 & 1 \\
    9 & -3 & -15 & -3 & -4 & -1 & 1 & 40 & -3 & -14 & -6 & 1 & 1 & -13 \\
    10 & 1 & -14 & -6 & -4 & -1 & 1 & 41 & 1 & -15 & -6 & -8 & 1 & -13 \\
    11 & -14 & -14 & -6 & -1 & -1 & 1 & 42 & 1 & -8 & -3 & -4 & -1 & -4 \\
    12 & 1 & -15 & -3 & -8 & 1 & -4 & 43 & -3 & -15 & -6 & 1 & 1 & -13 \\
    13 & -3 & -14 & -6 & 1 & 1 & 1 & 44 & 1 & -14 & -3 & -4 & -1 & -4 \\
    14 & -3 & -15 & -3 & -1 & -1 & 1 & 45 & -1 & -15 & -6 & -1 & -1 & 1\\
    15 & -1 & -8 & -6 & -4 & -1 & 1 & 46 & 1 & -15 & -3 & -8 & -10 & -13 \\
    16 & -1 & -15 & -3 & 1 & -1 & -13 & 47 & 1 & -15 & -6 & -4 & 1 & 1 \\
    17 & -1 & -15 & -3 & -1 & -1 & 1 & 48 & 1 & -10 & -6 & 1 & 1 & -13 \\
    18 & 1 & -14 & -3 & 1 & 1 & -4 & 49 & 1 & -10 & -3 & -10 & -10 & -13 \\
    19 & 1 & -15 & -6 & -4 & 1 & -13 & 50 & 1 & -15 & -3 & -4 & -10 & -13 \\
    20 & 1 & -14 & -6 & 1 & -1 & 1 & 51 & 1 & -15 & -3 & 1 & 1 & -13 \\
    21 & -3 & -15 & -6 & -4 & 1 & 1 & 52 & 1 & -15 & -3 & -10 & -10 & 1 \\
    22 & 1 & -14 & -3 & -4 & 1 & -4 & 53 & 1 & -10 & -3 & 1 & -10 & -13 \\
    23 & -3 & -15 & -6 & -1 & -1 & 1 & 54 & -1 & -15 & -3 & -4 & -1 & 1 \\
    24 & -3 & -15 & -6 & 1 & 1 & 1 & 55 & 1 & -14 & -6 & 1 & 1 & -13 \\
    25 & 1 & -15 & -3 & 1 & 1 & -4 & 56 & -1 & -8 & -6 & -1 & -1 & 1 \\
    26 & 1 & -14 & -3 & 1 & 1 & 1 & 57 & -14 & -15 & -3 & 1 & 1 & -4 \\
    27 & 1 & -15 & -3 & -4 & 1 & -13 & 58 & 1 & -15 & -6 & 1 & -1 & 1 \\
    28 & -1 & -15 & -6 & -4 & -1 & 1 & 59 & -3 & -15 & -3 & -4 & -10 & 1 \\
    29 & 1 & -10 & -6 & -4 & 1 & -13 & 60 & 1 & -15 & -3 & -8 & 1 & -13 \\
    30 & 1 & -15 & -6 & 1 & 1 & -13 & 61 & -1 & -14 & -6 & -1 & -1 & 1 \\
    31 & -1 & -14 & -3 & -4 & -1 & 1 & & & & & & &\\
  \bottomrule
  \end{tabular}
  \begin{tablenotes}
    \item The numbers represent different verification results as shown in Table~\ref{tab:classification}.
  \end{tablenotes}
 \end{threeparttable}}
\end{table*}

\begin{table}[t]
	\centering
	\caption{Discrepancies generated in first training episode: Amount}
	\label{tab:discrepancies-number}
	\begin{tabular}{ccc}
		\toprule
		\textbf{Certificate Size} & \textbf{Discrepancy Amount} & \makecell[bc]{\textbf{Proportion}\\(discrepancy/cert library)}\\
		\midrule
		6057& 3605& 59.5\%  \\
		15179& 9404& 61.9\%  \\
		84063& 49881& 59.3\%  \\
		148939& 78006& 52.3\% \\
		181900& 84661&46.5\% \\
		\bottomrule
	\end{tabular}
\end{table}

Table~\ref{tab:discrepancies-number} shows the discrepancy amount obtained by DRLgencert, and we could find the performance is quite efficient. Compared with Frankencert~\cite{brubaker2014using},  DRLgencert could get more discrepancy-triggering certificates. Using the 181,900 certificates as the training set, DRLgencert got 84,661 discrepancy-triggering certificates (46.5\% effectiveness) in first training episode of DRL network. Based on the same 181,900 certificates, Frankencert\footnote{We also tried to deploy Mucert~\cite{chen2015guided} and compared the results. However, Mucert was not maintained anymore, and the deployment procedure is not very friendly, which requires the customized execution environment and program instrumentation. After multiple attempts, we still did not run Mucert successfully.} only achieved 54,791 discrepancy-triggering certificates (30.1\% effectiveness) in a total of 181,900 outputted certificates.

\subsection{Flaws in Certificate Verification Code}
Based on the generated discrepancies, we found 23 flaws in these tested SSL/TLS implementations, as listed in Table~\ref{tab:flaw}. Among them, there are 14 flaws not discussed before. These flaws may lead to severe security threats, such as man-in-the-middle attacks. We have reported all of them to the corresponding vendors, and the assessments are in process. Here we analyze some typical flaws as showcases.

\begin{table*}[t]
	\centering
	\begin{threeparttable}
		\caption{Discovered flaws}
		\label{tab:flaw}
		\begin{tabular}{|l|c|c|c|c|c|c|}
			\toprule
			\textbf{Flaw Type}& \textbf{GnuTLS} & \textbf{MatrixSSL} & \textbf{MbedTLS} & ~~~\textbf{NSS}~~~ & \textbf{OpenSSL} & \textbf{wolfSSL} \\
			\midrule
			Incorrect checking of time& & 2 &1 & &  & \\
			\hline
			Incorrect checking of v1/v2 with X.509 v3 extension& 1 &1  & & 1& 1& \\
			\hline
			Incorrect checking of v1/v2 intermediate certificate& 1 &1  & & 1 & 1 & \\
			\hline
			Incorrect checking of version& 1 & & & 1& 1 & 1\\
			\hline
			Incorrect checking of serial number& 1 & & & 1& 1 & \\
			\hline
			Inappropriate error report& 1 &  1 & & 1& 1 & 1\\
			\midrule
			\textbf{Total} & \textbf{5} &\textbf{5} &\textbf{1} &\textbf{5} & \textbf{5}&\textbf{2}\\
			\bottomrule
		\end{tabular}
	\end{threeparttable}
\end{table*}


\subsubsection{Incorrect checking of time}
\label{section:Incorrect Checking of Time}
The content of the certificate contains two timestamps: notbefore and notafter. When the certificate is validated, the current time must be after notbefore, before notafter. Otherwise, the certificate is invalid. The current time must be GMT\footnote{Greenwich Mean Time, the mean solar time at the Royal Observatory in Greenwich, London.}. After analysis, we found that in these 6 SSL/TLS implementations, MatrixSSL and mbedTLS both use local system time instead of GMT.

Another finding is: according to RFC5280~\cite[4.1.2.5]{cooper2008internet}, the time checking should check whether notbefore is earlier than the current time and whether notafter is later than the current time. However, MatrixSSL is not this case. The relevant code of time checking in MatrixSSL is as follows (Listing 1):

\begin{lstlisting}[title={Listing 1: Time checking code in MatrixSSL (part 1)},basicstyle=\ttfamily\footnotesize]
if (psBrokenDownTimeCmp(&beforeTime, &timeNowLinger) > 0)
{
    /* beforeTime is in future. */
    cert->authFailFlags |= PS_CERT_AUTH_FAIL_DATE_FLAG;
}
else if (psBrokenDownTimeCmp(&timeNow, &afterTimeLinger) > 0)
{
    /* afterTime is in past. */
    cert->authFailFlags |= PS_CERT_AUTH_FAIL_DATE_FLAG;
}
\end{lstlisting}

In Listing 1, the ``timeNowLinger'' and ``afterTimeLinger'' variables are one day later than ``timeNow'' and ``afterTime''. The relevant code is shown in Listing 2.

\begin{lstlisting}[title={Listing 2: Time checking code in MatrixSSL (part 2)},basicstyle=\ttfamily\footnotesize]
/* The default value of allowed mismatch in times in X.509 messages and the local clock. The default value of 24 hours is mostly equivalent to old MatrixSSL behavior of ignoring hours, minutes and seconds in X.509 date comparison. Note: There is different value for CRL (PS_CRL_TIME_LINGER) and OCSP (PS_OCSP_TIME_LINGER). */
#  define PS_X509_TIME_LINGER (24 * 60 * 60)
...
memcpy(&timeNowLinger, &timeNow, sizeof timeNowLinger);
err = psBrokenDownTimeAdd(&timeNowLinger, PS_X509_TIME_LINGER);
\end{lstlisting}

As we can see from the above code, in MatrixSSL, the time check is unique. It compares notbefore with (current time + 1 day) and compares (notafter + 1 day) with the current time. So, if MatrixSSL is used to validate the certificate, the certificate will take effect one day in advance and postpone one day expired. In some scenarios, the lifetime of a certificate may be very short. For example, in Google's ALTS protocol~\cite{ALTS}, a handshake certificate usually is valid for 20 hours only. The error of extra one day before or after is undoubtedly a serious problem.

\subsubsection{Incorrect checking of v1/v2 with X.509 v3 extension}
\label{section:Incorrect checking of v1/v2 with x509 v3 extension}
According to discrepancy analysis, it is found that NSS, OpenSSL, and GnuTLS accept version 1 certificates with v3 extensions, wolfSSL rejects them. For version 2 certificates with v3 extensions, NSS, OpenSSL, GnuTLS, MatrixSSL accept them, and wolfSSL rejects them. According to the RFC documentation, \emph{this field (v3 extensions) MUST only appear if the version is 3}~\cite[4.1.2.9]{cooper2008internet}. So, every v1 or v2 certificate with v3 extensions is unqualified and should not be trusted. Our analysis found that one of the main reasons for this is that in the program code, version testing and extension testing are two independent content. The program does not judge whether the extension can exist based on the version.

\subsubsection{Incorrect checking of v1/v2 intermediate certificate}
\label{section:Incorrect checking of version 1/2 intermediate certificate}
Applications may locally trust a v1 root CA. So, v1 root certificate can be trusted. However, all version 1 and version 2 intermediate certificates must be rejected unless they can be verified to be a CA certificate through out-of-band.

If there is a trusted version 1 or version 2 intermediate certificate, since there is no v3 extension to limit its scope, they can act as a rough CA. They can sign any domain name, any number of next level certificates. These subordinate certificates can be used for man-in-the-middle attacks and other bad intentions. The related statements exist in the official RFC documentation: \emph{If certificate i is a version 1 or version 2 certificate, then the application MUST either verify that certificate i is a CA certificate through out-of-band means or reject the certificate. Conforming implementations may choose to reject all version 1 and version 2 intermediate certificates}~\cite[6.1.4(k)]{cooper2008internet}.

In our test, we found that: NSS, OpenSSL, GnuTLS, MatrixSSL all accept v1/v2 intermediate certificates and consider these certificates valid. Only wolfSSL rejected it. MbedTLS rejected these because of other unrelated reasons, mainly failing to load the certificate.

\subsubsection{Incorrect checking of version}
\label{section:Incorrect checking of version}
V3 certificates are now widely used and accepted, and in some cases, v1 and v2 certificates are considered valid. Certificates of any other versions are all invalid. However, the test found GnuTLS, NSS, OpenSSL, wolfSSL will accept v4 certificates, although v3 is the latest version of the X.509 certificate.

\subsubsection{Incorrect checking of serial number}
\label{section:Incorrect checking of serial number}
One of the requirements for the serial number in the RFC documentation is: \emph{the serial number MUST be a positive integer assigned by the CA to each certificate}~\cite[4.1.2.2]{cooper2008internet}. However, multiple certificates with negative sequence numbers were found in the generated discrepancy-triggering certificates, and they are trusted by GnuTLS, NSS, and OpenSSL.

Another requirement for serial numbers is: \emph{Conforming CAs MUST NOT use serialNumber values longer than 20 octets}~\cite[4.1.2.2]{cooper2008internet}. This statement is somewhat ambiguous. The first meaning is: CA certificate must not contain serialNumber values longer than 20 octets. The second meaning is: CAs must not assign serialNumber values longer than 20 octets to next level certificate. In discrepancy-triggering certificates, there are certificates containing a serial number with a length of 37 octets, and GnuTLS, NSS, and OpenSSL accepted the certificate. If the RFC documentation requirements refer to the second meaning or both, GnuTLS, NSS, and OpenSSL obviously violate the RFC documentation requirements.

\subsubsection{Inappropriate error report}
\label{section:Inappropriate error report}
In some cases, the same certificate may contain several different types of insecure contents that can result in varying degrees of security threats. The test found that in six programs, only mbedTLS gave multiple security warnings at the same time, and other programs reported only one security warning. The reason is that in the verification code of these programs when the first security threat is found, the verification code will directly return. This will introduce a security risk. If a certificate has many different security threats at the same time, for example, expiration for one week and self-signed, users will probably only receive a warning that the certificate expires one week. The bigger security threat self-signed is obscured. Some careless users, aware of the certificate expires only one week, likely choose to manually trust the certificate since it was just invalid, and this may result in a man-in-the-middle attack.

Inspired by this, we made a certificate, containing self-signed, expired, unstructured subject name. The unstructured subject name means the name of the subject in the certificate does not fit the document format, but this does not introduce a security problem. When verifying the certificate, we found that mbedTLS reported both self-signed and expired security warnings. OpenSSL and wolfSSL showed a self-signing warning. GnuTLS and MatrixSSL showed parsing errors (for the unstructured subject name), and we think this description is reasonable and sufficient since this implies the certificate has not been validated. However, NSS only displayed expiration warning. So from this perspective, the programs using NSS's service may suffer from the danger we previously assumed.

%% file: relatedwork.tex
\section{Related Work}
\label{sec:relatedwork}
There are already efforts of detecting vulnerabilities in SSL/TLS implementations. In this section, we will review some typical works. Also, some application cases of deep learning in security areas will be retrospected.  

\subsection{Security of SSL/TLS Implementations}

The SSL and TLS protocols are the foundation of modern network security. Whether SSL/TLS tools and libraries are correctly implemented are vital, and several previous works focused on validating their implementations.

Stone et al.~\cite{stone2017spinner} focused on the certificate hostname verification in TLS implementations. They showed that in security-sensitive applications, if the certificate pinning is used, it can hide the lack of proper hostname verification, enabling MITM attacks. They also presented a tool to detect such kind of vulnerability.

Sivakorn et al.~\cite{sivakorn2017hvlearn} directly tested the hostname verification in TLS implementations. They presented HVLearn, a novel black-box testing framework for analysis, which is based on automata learning algorithms. Moreover, this framework helps to find several unique violations of the RFC specifications in hostname verification implementations.

Acer et al.~\cite{acer2017wild} investigated the root causes of Chrome HTTPS certificate errors. Since hundreds of millions of spurious browser warnings are triggered per month, they frustrated users and undermined the trust in browser warnings, which may cause users to ignore the real warning. This finding shows that client-side or network issues caused more than half of errors. Based on this finding, Acer et al. redesigned the warnings of the browser. In our paper, we have a similar discovery, also about the wrong warning. We conducted a differential testing on multiple SSL certificate validation codes and found a phenomenon that multiple verification codes gave different error warnings for the same certificate, and each verification code gave only one error warning. So we refer that if there are a slight error and a serious error in a certificate, there may be a validation code that shows only minor error warning to users and masks serious error.

Hawanna et al.~\cite{hawanna2016risk} proposed a framework which assesses the risk associated with X.509 certificates. By using the random forest machine learning algorithm, this framework provides users with three risk levels -- high risk, medium risk, and low risk, and mentions due to which parameter it bears the risk. Another example of applying machine learning methods to HTTPS security is the work of Dong et al.~\cite{zheng}. Unlike the above work, they did not focus on the flaws in TLS implementations. They aimed to design a method for distinguishing rogue certificates which are valid certificates issued by a legitimate certificate authority (CA) but untrustworthy. In order to implement this method, they built machine-learning models with Deep Neural Networks (DNN) to automate classification. Similar to their idea, encouraged by the powerful performance of machine learning, we also apply machine learning to test SSL/TLS implementations. The difference is that we focus on the implementation errors and logic errors in the certificate validation phase. Also, we combine differential testing and deep reinforcement learning, which is different from the models used in the work of Hawanna et al. and Dong et al. and do not need to set the label for the data set.

The most closely work to ours is Frankencert~\cite{brubaker2014using} and Mucert~\cite{chen2015guided}. We use the same test method -- differential testing to test the SSL/TLS implementations. However, since both Frankencert and Mucert generate new certificates by random combinations of parts of certificates, introducing too much randomness. As a result, most of the generated certificates are invalid or valueless for differential testing. In contrast, we are using new ways to get a large number of certificates with the information learned in a neural network. The experimental results confirmed that we can get more valuable certificates.

\subsection{Applications of Deep Learning in Security}
There are many applications of deep learning in security. One of the primary applications is to develop classifiers for security tasks. Prade et al.~\cite{pradel2017deep} presented a general framework, called DeepBugs which extracts positive training examples from a code corpus, leverages simple program transformations to create negative training examples, and trains a model to distinguish these two. Tsai et al.~\cite{tsai2009intrusion} reviewed 55 related studies focusing on developing single, hybrid, and ensemble classifiers. The work of Dong et al. mentioned above is also a specific example of such kind of applications.

Godefroid et al.~\cite{godefroid2017learn} demonstrated using sample inputs and neural-network-based statistical machine-learning to automate the generation of an input grammar suitable for input fuzzing. The Portable Document Format (PDF)  was used as the complex input format in a case study, and they used recurrent neural network (RNN)~\cite{graves2009novel} to capture the structure of PDF and then generate new PDF files to fuzz PDF parser.

%% file: limitation.tex

%% file: conclusion.tex
\section{Conclusion}
\label{sec:conclusion}
In this paper, we presented a new idea of applying deep learning to differential testing of certificate validation. We introduce DRLgencert, a framework using deep reinforcement learning to generate discrepancy-triggering test cases for testing SSL/TLS implementations. The results show that DRLgencert is a valid and promising automatic system for generating test cases. After the first round of training, statistics show that it can modify certificates to discrepancy-triggering cases with a probability of more than 46\%. By conducting experiments in the real environment, DRLgencert successfully identified 23 certificate verification flaws on six popular SSL/TLS implementations.


%% file: acknowledge.tex
\section{Acknowledgements}
This work is partially supported by National Natural Science Foundation of China (91546203), the Key Science Technology Project of Shandong Province (2015GGX101046), the Shandong Provincial Natural Science Foundation (ZR2014FM020), Major Scientific and Technological Innovation Projects of Shandong Province, China (No. 2017CXGC0704), Fundamental Research Fund of Shandong Academy of Sciences (No.2018:12-16), and the Fundamental Research Funds for the Central Universities (No.21618330).